\documentclass[english,aps,twocolumn,prd,superscriptaddress,showpacs,preprintnumbers,amsmath,amssymb]{revtex4}

\usepackage{subfigure}
\usepackage{amsmath,amssymb}
\usepackage{graphicx}
\usepackage{rotating}
\usepackage{bm}
\usepackage{dsfont}
\usepackage{color}
\usepackage[dvips]{epsfig}

\def\di{\displaystyle}

\def\bg{\begin{eqnarray}\begin{array}{rcl}\displaystyle}
\def\eg{\end{array} &\di    &\di   \end{eqnarray}}
\def\bm#1{\begin{eqnarray}\begin{array}{#1}\di}
\def\bmo#1{\begin{eqnarray*}\begin{array}{#1}\di}
\def\bml#1#2{\begin{eqnarray}\begin{array}{#1}\label{#2}\di}
\def\bgo{\begin{eqnarray*}\begin{array}{rcl}\displaystyle}
\def\ego{\end{array} &\di    &\di \nonumber  \end{eqnarray*}}

\def\btensor#1#2{\renew\left#1\begin{array}{#2}\di}
\def\brtensor#1#2#3{\ren#3\left#1\begin{array}{#2}}
\def\botensor#1#2{\renew\left#1\begin{array}{#2}}
\def\etensor#1{\end{array}\right#1}

\def\eq#1{(\ref{#1})}
\def\Eq#1{Eq.~(\ref{#1})}

\def\id{1\!\mbox{l}}

\def\s0#1#2{\mbox{\small{$ \frac{#1}{#2} $}}}
\def\0#1#2{\frac{#1}{#2}}

\def\dr{{D\!\llap{/}}\,}

\def\CO{{\mathcal O}}
\def\CP{{\mathcal P}}

\def\CZ{{\mathcal Z}}

\def\Z{\mathds{Z}}
\def\ren#1{\renewcommand{\arraystretch}{#1}}

\def\renew{\renewcommand{\arraystretch}{1}}
\definecolor{blue}{rgb}{0,0,1}

\definecolor{green}{rgb}{0,1,0}

\definecolor{red}{rgb}{1,0,0}

\newcommand{\Tr}{\mathrm{Tr}}

\newcommand{\tr}{\mathrm{tr}}
\newcommand{\E}{\mathrm{e}}
\newcommand{\I}{\mathrm{i}}
\newcommand{\be}{\begin{eqnarray}}
\newcommand{\ee}{\end{eqnarray}}

\newcommand{\Nc}{N_{\text{c}}}

\usepackage{babel}
\makeatother

\begin{document}

\title{On the relation of quark confinement and chiral symmetry breaking}
\pacs{05.10.Cc,11.10.Wx,12.38.Aw}
%\date{\today}                                           
\author{Jens~Braun}
\affiliation{Theoretisch-Physikalisches Institut, Friedrich-Schiller-Universitat Jena, 
Max-Wien-Platz 1, 07743 Jena, Germany}
\author{Lisa M.~Haas} 
\affiliation{Institut f\"ur Theoretische Physik, University of Heidelberg, 
Philosophenweg 16, 62910 Heidelberg, Germany}
\affiliation{ExtreMe Matter Institute EMMI, GSI Helmholtzzentrum f\"ur 
Schwerionenforschung, Planckstr. 1, 64291 Darmstadt, Germany.}
\author{Florian~Marhauser}
\affiliation{ExtreMe Matter Institute EMMI, GSI Helmholtzzentrum f\"ur 
Schwerionenforschung, Planckstr. 1, 64291 Darmstadt, Germany.}
\affiliation{Institut f\"ur Theoretische Physik, University of Heidelberg, 
Philosophenweg 16, 62910 Heidelberg, Germany}
\author{Jan M. Pawlowski}
\affiliation{Institut f\"ur Theoretische Physik, University of Heidelberg, 
Philosophenweg 16, 62910 Heidelberg, Germany}
\affiliation{ExtreMe Matter Institute EMMI, GSI Helmholtzzentrum f\"ur 
Schwerionenforschung, Planckstr. 1, 64291 Darmstadt, Germany.}

\begin{abstract}
  We study the phase diagram of two flavour QCD at imaginary chemical
  potential in the chiral limit. To this end we compute order parameters for chiral
  symmetry breaking and quark confinement. The interrelation of quark
  confinement and chiral symmetry breaking is analysed with a new
  order parameter for the confinement phase transition. We show that
  it is directly related to both, the quark density as well as the
  Polyakov loop expectation value. Our analytical and numerical results
  suggest a close relation between the chiral and the confinement
  phase transition.
\end{abstract}

\maketitle

{\it Introduction -} Quantum Chromodynamics (QCD) at finite
temperature and density is a very active area of research. The
equation of state of QCD and, in particular, the nature of the
transition from the hadronic phase with broken chiral symmetry
to the chirally symmetric deconfined quark-gluon plasma phase is 
of great importance for a better understanding of the
experimental data, e.g.~\cite{BraunMunzinger:2003zd}.

For full QCD with dynamical quarks one expects the confinement phase
transition to be a crossover as quarks explicitly break the
underlying center symmetry of the gauge group. The nature of the
chiral phase transition primarily depends on the value of the current
quark mass, which explicitly breaks chiral symmetry, as well as on the
strength of the chiral anomaly~\cite{Pisarski:1983ms}.  While the
confinement phase transition is driven by gluodynamics, the chiral
phase transition is governed by strong quark interactions. Hence, it
is a highly non-trivial observation that both lie remarkably close at
least for small quark chemical
potentials~\cite{Aoki:2006br,de Forcrand:2002ci}. An understanding of
this interrelation is subject of an ongoing debate.

The chiral properties of the theory are accessible through the
spectrum of the Dirac operator \cite{Banks:1979yr}. Recently it has
been shown that also the confining properties of the theory can be
accessed through its spectrum~\cite{Gattringer:2006ci,%
  Synatschke:2007bz,Bruckmann:2008sy,Fischer:2009wc}.  In this Letter
we analyse the deconfinement and chiral phase transition and their
interrelation both analytically and numerically for two flavour QCD in
the chiral limit. The interrelation is studied with the help of a new order
parameter for the confinement phase transition which is related to the
quark density.  The numerical analysis is performed
with functional renormalization group (RG) methods, for reviews
see~\cite{Litim:1998nf,Pawlowski:2005xe}, and suggests that the
deconfinement phase transition is indeed correlated with
chiral symmetry restoration.\\[-2ex]

{\it Order parameters -} The Polyakov loop variable $L(\vec{x})$, 
\begin{equation}\label{eq:Polloop}
L(\vec x)=\frac{1}{\Nc} \tr\, \CP(\vec x)\;\text{with}\;
{\cal P}(\vec x)  =\text{P}\, \E ^{ \I g\int_0^\beta d t\, {A}_0(t,\vec x)}\,,
\end{equation}
in QCD with $N_c$ colors and infinitely heavy quarks is related to the
operator that generates a static quark~\cite{Polyakov:1978vu}. In
\eq{eq:Polloop} the trace is evaluated in the fundamental
representation, and $\text{P}$ stands for path ordering. We can
interpret the logarithm of the expectation value $\langle L\rangle$ as
half of the free energy $F_{q\bar q}$ of a static quark--anti-quark
pair at infinite distance. Moreover the expectation value $\langle
L\rangle$ is an order parameter for center symmetry of the gauge
group, e.g.~\cite{Greensite:2003bk}. To see this, we consider
gauge transformations $U_z(t,x)$ with $U_z^{-1}(0,\vec x) U_z(\beta,
\vec x) =z$, where $z\in \CZ$ is an element of the center $\CZ$ of the
gauge group. Under such a transformation the Polyakov loop is
multiplied with a center element $z$, $L(\vec x)\to z\, L(\vec
x)$. Hence a center-symmetric confining disordered ground state with
$F_q \to\infty$ is ensured by $\langle L\rangle=0$. In turn,
deconfinement with $F_q <\infty$ is signaled by $\langle L\rangle\neq
0$. This implies center-symmetry breaking in the ordered phase.

It follows immediately that {\it any} observable, which transforms
non-trivially under center transformations, serves as an order
parameter. This has been exploited in
\cite{Gattringer:2006ci,Synatschke:2007bz,%
  Bruckmann:2008sy,Fischer:2009wc}, where the spectral properties of
the Dirac operator have been related to the expectation value of the
Polyakov loop. The relation stems from the observation that, in
contradistinction to the gauge fields, the periodicity properties of
the quark fields change under application of a gauge transformation
$U_z$,
\begin{equation}\label{eq:psiU}
  \psi^{U_z}(t+\beta, \vec x) = - z \psi^{U_z}(t, \vec x)\,,
\quad {\rm with}\quad \beta=\01T\,. 
\end{equation}
A straightforward generalisation of the boundary conditions of the
quarks  yields
\begin{equation}\label{eq:psiboundary}
  \psi_\theta (t+\beta, \vec x) = - e^{2 \pi i \theta } \psi_ \theta(t, \vec x)\,. 
\end{equation}
This includes \eq{eq:psiU} with the center phases $z=\id\, e^{2 \pi
  i\theta_z }$, e.g.  $\theta_z=0, 1/2$ in $SU(2)$, and
$\theta_z=0,1/3, 2/3$ in $SU(3)$. Quarks with the boundary
conditions \eq{eq:psiboundary} can be rewritten in terms of quarks 
with physical anti-periodic boundary conditions,
\begin{equation}\label{eq:psitheta}
  \psi_\theta(x) = e^{2 \pi \theta i\,t / \beta} \psi(x)\,\quad {\rm with}
  \quad \psi(x)=\psi_{\theta=0}(x)\,.
\end{equation} 
Due to the periodicity in $\theta$, general observables
$\CO_\theta=\langle O[\psi_\theta]\rangle$ can be represented in a
Fourier decomposition,
\begin{equation}\label{eq:fourier}
 \CO_{\theta}=\sum_{l\in \Z} e^{2 \pi i l\theta} O_l\,. 
\end{equation}
This implies that the {\it dual} observables $O_l$ change under a
center sensitive gauge transformation $U_z$ into $z^l O_l$. Hence
every moment $O_{l}$ with $l\in \Z$ and $l\ \text{mod}\;N_c\neq 0$ has
to vanish in the center symmetric phase as it is proportional to a
sum over center elements $z$,
\begin{equation}\label{eq:center0}
\sum_{z\in \CZ} z^l=N_c\,\delta_{l\!\!\!\!\mod N_c,0}\,.
\end{equation}
Note that ${\cal L}_\theta= e^{2 \pi i\,\theta} \langle L\rangle$
reflects the boundary conditions \eq{eq:psiboundary} and fits into the
definition of dual observables. The observables ${\cal O}_\theta$ can
either be evaluated in QCD with anti-periodic quarks
\cite{Gattringer:2006ci,Synatschke:2007bz,%
  Bruckmann:2008sy,Fischer:2009wc}, or in QCD${}^{\ }_\theta$ with
quarks having $\theta$-dependent boundary conditions.

In summary, the moments $O_l$ with $l\mod N_c\neq 0$ are order
parameters for the confinement phase transition in QCD if evaluated in
both, QCD and QCD${}_\theta$. In particular the first moment $O_{1}$
is an order parameter for all $N_c$,
\begin{equation}\label{eq:conforder}
\tilde \CO=\int_0^1 d\theta\, e^{-2 \pi i \theta}  \CO_\theta\,. 
\end{equation}
For example, the dual Polyakov loop in QCD is $\langle L\rangle$. 
\\[-2ex]

{\it QCD at imaginary chemical potential -} In the present work we
mainly concentrate on QCD${}^{\ }_\theta$. Its generating functional is
\begin{equation}\label{eq:Zqcd}
  Z_\theta[J]\!=\! 
  \int dA \, d \psi_\theta d\bar\psi_\theta\,
  \text{e}^{- S[A,\psi_\theta,\bar\psi_\theta] + \int J\phi_\theta}\,,
\end{equation}
with $\phi_\theta=(A,\psi_\theta,\bar\psi_\theta,...)$ and
$J=(J_A,\bar\eta, \eta,...)$. The dots stand for the ghost fields and
composite hadronic fields, see
e.g.~\cite{Litim:1998nf,Pawlowski:2005xe}. Here, $S$ denotes the
standard QCD action and includes a Dirac action with
$\theta$-dependent quark fields. With \eq{eq:psitheta} we have
\begin{equation}\label{eq:Sdiractheta}
\int \bar\psi_\theta \left( i \dr+m \right)\psi_\theta=\int \bar\psi\left( i \dr +m -
2 \pi \01\beta \gamma_0 \theta\right)\psi\,, 
\end{equation}
where $\dr=(\partial\!\!\!\slash +g A\!\!\!\slash)$. The rhs of
\eq{eq:Sdiractheta} is nothing but the Dirac action with an imaginary
chemical potential $\mu=2 \pi i\,\theta/\beta$.  If $\theta$ takes one
of the center values $\theta_z$, we can define
$\psi^{U_z}=\psi_{\theta_z}^{\ }$ with $U_z$ as in \eq{eq:psiU} and
anti-periodic $\psi$. We conclude that center phases $\theta_z$ can be
absorbed in center transformations of the gauge field, $A\to
A^{U^\dagger_z}$. The generating functional at vanishing current $J$
has the Roberge-Weiss (RW) periodicity \cite{Roberge:1986mm}, see also
\cite{Kratochvila:2006jx,de Forcrand:2002ci,Sakai:2008py},
\begin{equation}\label{eq:thetaperiodic}
Z_\theta[0]=Z_{\theta+1/N_c}[0]\,.  
\end{equation} 
Furthermore, observables ${\cal O}_\theta$ in QCD${}_\theta^{\ }$ are
RW-symmetric for $J=0$. Hence only the center-symmetric Fourier
coefficients $O_{N_c l}$ are non-vanishing.  In turn, a non-vanishing
current $J_A$ for the gauge field breaks the RW-symmetry, and leads to
$O_l\neq 0$ for $l \mod N_c \neq 0$.  This leads us to a simple and
easily accessible confinement order parameter in QCD${}_\theta^{\ }$,
the dual density:
\begin{equation}\label{eq:dualdensity}
  \tilde{n}[\phi_J]:=\int_0^{1} d\theta\, e^{-2 \pi i \theta} n_\theta
  =\I\beta \int_0^{1} d\theta\, e^{-2 \pi i \theta} \ln Z_\theta[J] \,, 
\end{equation}  
with $\phi_J=\langle \phi\rangle_J$. The density $n_{\theta}$ is
the derivative of the partition function w.r.t.\ the chemical potential $2 \pi\,\theta/\beta$,
\begin{equation}
  n_\theta[\phi_J] =\int d^4 x\,\langle  \bar{\psi}
  \gamma _0 \psi \rangle _{\theta}=\frac{\beta}{2\pi}
  \partial _{\theta}\ln Z_\theta[J]\,.
\end{equation}
On the rhs of \eq{eq:dualdensity} we have integrated by parts and made
use of $Z_0[J]=Z_{1}[J]$. The dual density $\tilde n[\phi]$ is
proportional to the first moment of the grand canonical potential in
the presence of a gauge field background $\varphi$. Hence, it grows
like $T^3$ at high temperatures, as the integrated $\theta$-dependence
is expected to be leading order.

\begin{figure}[]
\includegraphics[clip,scale=0.6]{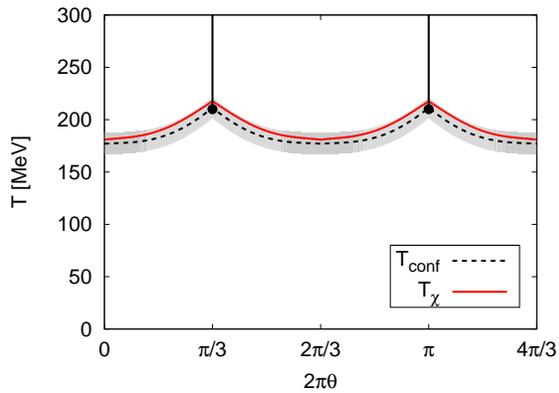}
\caption{Chiral ($T_{\chi}$) and confinement temperature
  ($T_{\text{conf}}$) as functions of temperature and boundary angle
  $\theta$. The width of $\chi_L=\partial_T L$ is displayed as a
  shaded area. The dots indicate the endpoints of the confinement RW-phase transitions.}
\label{fig:phasetheta} 
\end{figure}

The above analysis for the dual density extends to general observables
$\CO_{\theta}[\phi_J]$. They constitute observables in {\it different}
theories distinguished by the boundary condition. Hence the order
parameters $\tilde{\cal O}$ in \eq{eq:conforder} in the presence of a
fixed background $\phi_J=\langle \phi\rangle_J$ only vanish if
QCD${}^{\ }_\theta$ is in the center symmetric phase for all boundary
conditions. However, the transition temperature in QCD, $T_{\rm
  conf}=T_{\rm conf}(\theta=0)$, is a lower bound for $T_{\rm
  conf}(\theta)$, see Fig.~\ref{fig:phasetheta}. Thus, the dual phase
transition temperatures $\tilde T_{\rm conf}$ are identical with the
physical one, $\tilde T_{\rm conf}= T_{\rm conf}$. \\[-2ex]

{\it Analytic properties -} Some interesting properties of dual order
parameters can be accessed analytically. To that end it is convenient
to study observables in terms of the quantum effective action
$\Gamma[\phi]= \int J \phi- \ln Z_\theta[J]$. The current $J$ is given
by $J=\delta\Gamma[\phi]/\delta\phi$, and vanishes on the equations of
motion with $\theta$-dependent mean value
$\bar\phi_\theta=\phi_{J=0}$. For example, the dual density
\eq{eq:dualdensity} turns into
\begin{equation}\label{eq:dualdensityG}
\tilde n[\phi]=- i\beta
\int_0^{1} d\theta\, e^{-2 \pi i \theta}\Gamma[\phi]\,.
\end{equation} 
As discussed above, we have $\tilde n[\bar\phi_\theta]=0$.  This
follows from the RW-symmetry of $\Gamma[\bar\phi_\theta]$ which is a
consequence of that of $Z_\theta$.  For its direct proof it is
sufficient to examine $\Gamma[\phi]$ for constant gauge field
configurations $A_0$ in the Cartan subalgebra. For example, for
$N_c=3$ we have Cartan fields $\beta g A_0=\varphi_3 \tau^3/2
+\varphi_8 \tau^8/2$ with Gell-Mann matrices $\tau^3,\tau^8$. The
$\theta$-dependence of the effective action $\Gamma[\phi]$ originates
in sums of terms with fermionic Matsubara frequencies stemming from
$D_0+2 \pi T\theta$: $2 \pi T (n+1/2+\theta+\beta g A_0/(2
\pi))$. Most of the $\theta$-dependence can be reabsorbed in a
$\theta$-dependent gauge field $A_0(\theta)$. For $N_c=3$ this leads
to Cartan components $\hat\varphi_3=\varphi_3-3 (2\pi) \theta$ and
$\hat\varphi_8=\varphi_8-\sqrt{3} (2\pi) \theta$ of
$A_0(\theta)$. Now, the RW-symmetry is explicit in the Matsubara
frequencies
\begin{equation}\label{eq:mats}
  2 \pi T\left(n+\frac{1}{2} +\0{1}{4 \pi}\Phi_i+ N_c 
    \delta_{i1} \theta\right)\,,\quad   i=1,...,N_c\,,
\end{equation}
where the $\Phi_i$'s are the eigenvalues of the matrix $2 \beta g
A_0(\theta)$: $\theta\to \theta+\theta_z$ is absorbed in a center
gauge transformation of the $\varphi_i$'s as well as in a shift of the
Matsubara sum.  Under this combined transformation the
$\hat\varphi_i$'s are invariant and so is the effective action. In
particular we conclude that any expansion scheme based on fixed field
variables $\hat\varphi_i$ is form-invariant under
$\theta\to\theta+\theta_z$. Moreover, the observables ${\cal
  O}_\theta[\bar\phi_\theta]$ are invariant, and hence $\tilde{\cal
  O}[\bar\phi_\theta]\equiv 0$. In turn, observables $\tilde
\CO[\phi]$ with $\theta$-independent gauge field background $\varphi$ are order
parameters for confinement as such a background explicitly breaks the
RW-symmetry. In particular this includes $\tilde {\cal O}[\phi]$
with $\varphi=\bar\varphi=\bar\varphi_{\theta=0}$ and $\varphi=0$. 

Simple observables $\tilde {\cal O}[\phi]$ follow directly from the
vertices $\Gamma^{(n)}[\phi]$ in QCD${}^{\ }_{\theta}$. This
includes the dual density \eq{eq:dualdensityG} as well as the dual
chiral condensate with ${\cal O_\theta}[\phi_J]=\int d^4 x\,
\langle \bar\psi_\theta\psi_\theta\rangle_J$ for either 
$\varphi_J=\bar\varphi_{\theta=0}$ and $\varphi_J=0$. The
first case with $\bar\varphi$ relates to the lattice computations in
QCD of dual order parameters
\cite{Gattringer:2006ci,Synatschke:2007bz,Bruckmann:2008sy}. The
latter choice has been used implicitly in \cite{Fischer:2009wc,Fischer}. An
even simpler observable is the dual quark mass parameter $\tilde {\cal
  M}$ with ${\cal M}_\theta[\phi]\sim\tr\,
\Gamma_{\bar{\psi}\psi}^{(2)}[\phi](p=0)$. The specific choice 
${\cal M}_\theta[\bar\phi_\theta]$ is directly related to the pion
decay constant $f_\pi$ in QCD${}_\theta^{\ }$.  A further prominent
example is the modified Polyakov loop variable $L_\theta=e^{2 \pi
  i\,\theta} L$,
\begin{equation}\label{eq:Ltheta}
  L_\theta[\varphi]=\0{1}{N_c} \sum_{i=1}^{N_c} \text{e}^{2 \pi i
    \left( \0{1}{4 \pi}\Phi_i[\hat\varphi]+ N_c \delta_{i1} 
      \theta\right)}\,, 
\end{equation}
with ${\cal L}_\theta=\langle L_\theta\rangle$. \Eq{eq:Ltheta} is
invariant under $\theta\to\theta+\theta_z$ at fixed $\hat\varphi$, and
hence $\tilde {\cal L}[\bar\phi_\theta]=0$. However, $\tilde
L[\bar\varphi]=L[\bar\varphi]$ simply is the Polyakov loop
variable introduced in~\cite{Braun:2007bx,Marhauser:2008fz} as an
order parameter for confinement. 
%
%%%%%%%%%%%%%%%%%%%%%%%%%%%%%%%%%%%%%%%%%%%%%%%%%%%%%%%%%%%%%%%%%%%%%
\begin{figure}[t!]
\centerline{\epsfig{file=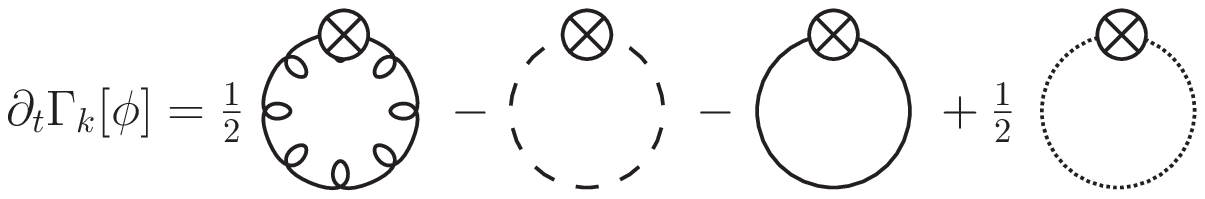,width=8cm}}
\caption{Functional flow for the effective action: The lines denote full field dependent 
propagators. Crosses denote the cut-off insertion $\partial_t R$.  }
\label{fig:funflow}
\end{figure}
%%%%%%%%%%%%%%%%%%%%%%%%%%%%%%%%%%%%%%%%%%%%%%%%%%%%%%%%%%%%%%%%%%%%%%%
%

The representation of the Polyakov loop in \eq{eq:Ltheta} leads to an
interesting observation: in phase-quenched QCD${}_\theta$ we are left with
the explicit $\theta$-dependence in the Matsubara frequencies
\eq{eq:mats}. Thus, any observable $\tilde {\cal O}$ in \eq{eq:conforder} obeys
\begin{equation}\label{eq:varphi-dep}
  \tilde \CO[\phi]=\int_0^1 d\theta
  \, e^{-2 \pi i \theta } \CO_\theta[0]\, L[\varphi]=
  \tilde \CO[0]\, L[\varphi] \,,  
\end{equation} 
for $\theta$-independent gauge field background $\varphi$ and
vanishing quark and mesonic backgrounds. In fully dynamical
QCD${}_\theta^{\ }$ the factorisation \eq{eq:varphi-dep} only holds approximately.\\[-2ex]

{\it QCD with functional methods -} Our numerical computations are
performed within the functional RG approach to the
quantum effective action $\Gamma_k$. Here, $k$ is an infrared cut-off
scale below which quantum fluctuations are suppressed. For $k\to0$ we
regain the full quantum effective action $\Gamma$.  The effective
action $\Gamma_k$ obeys the Wetterich equation~\cite{Wetterich:1993yh},
\begin{equation}\label{eq:FRG} 
  \partial_t  \Gamma _k [\phi] =\frac{1}{2} \Tr \0{1}{
    \Gamma_k^{(2)}[\phi]+R_k} \partial_t R_k \,, 
\end{equation} 
with $t=\ln k/\Lambda$ and cut-off functions $R_k$ that provide
infrared cut-offs for all fields, for reviews on gauge theories
see~\cite{Litim:1998nf,Pawlowski:2005xe}.  The diagrammatic
representation of \eq{eq:FRG} is provided in Fig.~\ref{fig:funflow}.
The RG flow for the dual density follows from \eq{eq:dualdensityG} and
is directly related to the flow of the effective action.  This is an
important property as the flow for $\Gamma_k$ is least sensitive to
the approximations involved. It also guarantees the maximal
disentanglement of the different field sectors, see
Fig.~\ref{fig:funflow}. We conclude that the dual order parameters, in
particular $\tilde n$, are dominated by the quark loop. Moreover, in
the present approach with dynamical mesonic degrees of freedom the
quark propagator is in leading order only sensitive to the chiral
properties, see e.g.~\cite{Gies:2002hq,Braun:2008pi}. Consequently,
the confinement temperature in QCD, derived from the dual density
$\tilde n$, has to agree approximately with the chiral critical
temperature. Thus the chiral and confinement phase transitions are
necessarily closely related. This
observation is sustained by our explicit computations, see below.\\[-2ex]
%
%%%%%%%%%%%%%%%%%%%%%%%%%%%%%%%%%%%%%%%%%%%%%%%%%%%%%%%%%%%%%%%%%%%%%%%%%%%%%%%
\begin{figure}[t!]
\includegraphics[clip,scale=0.6]{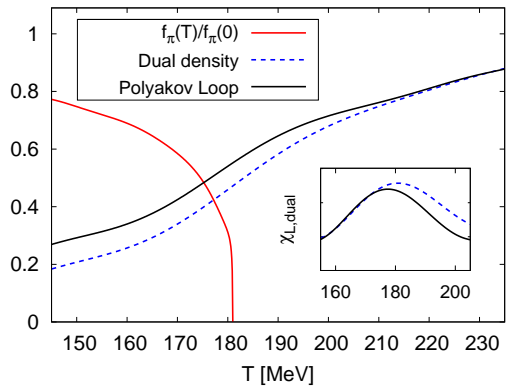}
\caption{The pion decay constant $f_\pi(T)/f_\pi(0)$, the dual density
  $\tilde n(T)/\tilde n(\infty)$, and the Polyakov loop $L[\bar
  \varphi](T)$ as functions of temperature, $\chi^{\ }_L=\partial^{\
  }_T L$, $\chi^{\ }_{\rm dual}=\partial^{\ }_T \tilde n$.}
\label{fig:orderparameters} 
\end{figure}
%%%%%%%%%%%%%%%%%%%%%%%%%%%%%%%%%%%%%%%%%%%%%%%%%%%%%%%%%%%%%%%%%%%%%%%%%%%%%%%

{\it Numerical results \& summary -} For our study of two flavour QCD
in the chiral limit, we solve the flow equation for the effective
action $\Gamma$ by combining results for the Yang-Mills part of QCD
\cite{Braun:2007bx,Fischer:2008uz}, as well as the matter part
\cite{Gies:2002hq,Braun:2008pi,Berges:2000ew}. The two sectors are
coupled by the dynamical quark-gluon interaction. This setting
incorporates the confining properties of QCD \cite{Braun:2007bx} via
the full momentum dependence of gluon and ghost propagators
\cite{Fischer:2008uz}.  The results for pure Yang-Mills agree
quantitatively with the corresponding lattice results. In the matter
sector mesonic degrees of freedom are dynamically included
\cite{Gies:2002hq,Braun:2008pi,Berges:2000ew}. Such a treatment of the
matter sector already provides quantitatively reliable results for the
meson spectrum, see e.g.\ \cite{Berges:2000ew}. It has been also
successfully implemented for the phase diagram of one flavour QCD at
finite chemical potential \cite{Braun:2008pi}.

In Fig.~\ref{fig:orderparameters} the temperature dependence of two
order parameters for confinement are shown, namely the Polyakov loop
variable $L[\bar\varphi]$ and the dual density $\tilde
n[\bar\phi]$. The crossover temperature $T_{\rm conf}$ is determined
by the peaks in the respective $T$-derivatives $\chi_L$ and $\chi_{\rm
  dual}$. Interestingly the factorisation \eq{eq:varphi-dep} works
quantitatively for the dual density in the full theory: $\tilde
n[\bar\phi]/\tilde n[0]$ and $L[\bar\varphi]$ agree on the percent
level. We have checked further order parameters such as the dual pion
decay constant. We find that the crossover temperatures extracted from
the dual density, the dual pion decay constant and the Polyakov loop
agree within a few MeV: $T_{\text{dual}}\approx T_{\text{conf}}\approx
178\,\text{MeV}$. This provides further support for the quantitative
reliability of the present approximation.
%%%%%%%%%%%%%%%%%%%%%%%%%%%%%%%%%%%%%%%%%%%%%%%%%%%%%%%%%%%%%%%%%%%%%%%%%%%%%%%
%%%%%%%%%%%%%%%%%%%%%%%%%%%%%%%%%%%%%%%%%%%%%%%%%%%%%%%%%%%%%%%%%%%%%%%%%%%%%%%

In Fig.~\ref{fig:orderparameters} we also show the pion decay constant
$f_\pi$.  It is proportional to the quark mass parameter ${\cal
  M}_\theta$ evaluated at $\bar\phi_\theta$, and is an order
parameter for the chiral phase transition. For $T\to 0$, $f_{\pi}$
approaches $90\,\text{MeV}$.  For $T>T_{\chi}\approx 181\,\text{MeV}$
the pion decay constant tends to zero and chiral symmetry is
restored. We observe a second order phase transition, and the critical
exponents such as $\nu$ signal the $O(4)$-universality class.  Most
importantly, the chiral phase transition and the confinement crossover
temperature agree at vanishing chemical potential.

An evaluation of the dual chiral condensate and the dual quark mass
parameter in QCD${}_\theta$ for vanishing gauge field background
$\varphi=0$ has been implicitly performed in \cite{Fischer:2009wc} and
\cite{Fischer} respectively. Evaluated at both, $0$ and $\bar\varphi$,
we find the expected periodicity of $\tilde M$ in $\theta\to
\theta\!+\!1$, and no RW-symmetry. For $\theta=1/2$ it can be shown
analytically that it grows with $T^{1/2}$ for large $T$. In turn, for
$\theta=0$ and $\varphi=\bar\varphi_{\theta=0}$ it agrees with $f_\pi$
and vanishes for large $T$, see Fig.~\ref{fig:orderparameters}. Below
the chiral phase transition temperature $T_{\chi}$ the mass parameter
$\tilde {\cal M}$ is a smooth function of $\theta$. However, a box-like
behavior emerges above $T_{\chi}$, see also \cite{Fischer}. Details
will be presented elsewhere. Here we simply note, that ${\cal
  M}_\theta[\phi]$ is an expansion parameter of the effective action
which only signals chiral symmetry breaking for
$\phi\!=\!\bar\phi_\theta$.

In Fig.~\ref{fig:phasetheta} we show the phase diagram of
QCD${}_\theta^{\ }$. The confinement and the chiral temperatures lie
close to each other for all imaginary chemical potentials. Their value
at $\theta=1/6$ is the endpoint $(T_{\text{RW}},\theta_{\text{RW}})
\approx (210 {\rm MeV},\,1/6)$ of the corresponding RW phase
transitions shown as a vertical line at $\theta=1/6$ in
Fig.~\ref{fig:phasetheta}.  Our results compare well to the lattice
results \cite{Kratochvila:2006jx}.  In the PNJL-model
\cite{Fukushima:2003fw} the lattice results have been reproduced by
adjusting model parameters connected to an eight quark interaction
\cite{Sakai:2008py}. In our approach to full QCD${}_\theta$ coinciding
temperatures result from the interplay of quantum fluctuations and are
not adjusted by hand. An estimate of the corresponding quantum
fluctuation within a Polyakov--quark-meson model also leads to
coinciding critical temperatures at real chemical potential
\cite{Schaefer:2007pw}. These results suggest that the differences
between $T_{\rm conf}$ and $T_{\chi}$ at both, real and imaginary
chemical potential, are mainly due to mean field or large $N_c$
approximations. The relevance of this observation for the quarkyonic
phase proposed in \cite{McLerran:2007qj} will be discussed elsewhere.

In summary our study suggests that the confinement and chiral
critical temperatures $T_{\rm conf}$ and $T_{\chi}$ are dynamically related and
agree within the error bars. At present, we extend our work to real
chemical potential.  This may help to shed some light on the current
debate concerning lattice simulations at finite chemical
potential.\\[-2ex]
 
{\it Acknowledgments -} We thank C.~S.~Fischer, C.~Gattringer, H.~Gies,
E.~Laermann, A.~Maas and A.~Wipf for discussions. This work is
supported by Helmholtz Alliance HA216/EMMI.

\end{document}